\documentclass[11pt]{article}

\usepackage{orcidlink,thumbpdf,lmodern}

\usepackage{Sweave}
\usepackage{amsmath}
\usepackage{ulem}
\usepackage{tikz}

\usepackage{framed}
\usepackage{bm} 
\usepackage{enumitem} 
\setlist[itemize]{noitemsep} 
\setlist[enumerate]{noitemsep} 
\usepackage{dsfont} 
\usepackage{amsfonts} 
\usepackage{arydshln} 
\usepackage{crossreftools} 
\usepackage{comment}
\usepackage[round]{natbib}
\usepackage{geometry}
\usepackage{pdfpages}
\usepackage{setspace}
\usepackage{authblk}

\newcommand{\proglang}[1]{\textsf{#1}}
\newcommand{\pkg}[1]{\textbf{#1}}
\newcommand{\code}[1]{\texttt{#1}}

\newcommand{\Yi}{Y_i}

\renewcommand{\Xi}{X_i}
\newcommand{\Xj}{X_j}
\newcommand{\Xk}{X_k}
\newcommand{\Di}{D_i}

\newcommand{\Zi}{Z_i}

\newcommand{\Zj}{Z_j}
\newcommand{\Zk}{Z_k}

\newcommand{\deltai}{\delta_i}
\newcommand{\epsi}{\epsilon_i}
\newcommand{\VOmega}{\mathcal{T}}
\newcommand{\Y}{\mathbf{Y}}
\newcommand{\D}{\mathbf{D}}
\newcommand{\hbold}{\mathbf{h}}

\newcommand{\deltabold}{\bm{\delta}}
\newcommand{\phibold}{\bm{\phi}}
\newcommand{\epsbold}{\bm{\epsilon}}

\newcommand{\Omegabold}{\mathbf{\Omega}}
\newcommand{\Rcal}{\mathcal{R}}

\newcommand{\Acalone}{\mathcal{A}_1}
\newcommand{\Acaltwo}{\mathcal{A}_2}

\newcommand{\hatfAonebold}{\hat{\mathbf{f}}_{\Acalone}}
\newcommand{\hboldAone}{\mathbf{h}_{\Acalone}}
\newcommand{\phiboldAone}{\phibold_{\Acalone}}
\newcommand{\epsboldAone}{\epsbold_{\Acalone}}

\newcommand{\DAonebold}{\D_{\Acalone}}
\newcommand{\DAtwobold}{\D_{\Acaltwo}}
\newcommand{\YAonebold}{\Y_{\Acalone}}

\newcommand{\Qmax}{Q_{\text{max}}}
\newcommand{\qcomp}{q_{\text{comp}}}

\newcommand{\betahat}{\hat{\beta}}
\newcommand{\basish}{\overset{\rightarrow}{\mathbf{v}}}
\newcommand{\basisphi}{\overset{\rightarrow}{\mathbf{w}}}

\newcommand{\Vcal}{\mathcal{V}}
\newcommand{\Vbold}{\mathbf{V}}
\newcommand{\Wbold}{\mathbf{W}}
\newcommand{\VboldAone}{\mathbf{V}_{\Acalone}}
\newcommand{\WboldAone}{\mathbf{W}_{\Acalone}}
\newcommand{\hatV}{\widehat{\Vbold}}
\newcommand{\hatW}{\widehat{\Wbold}}
\newcommand{\hatVAone}{\widehat{\Vbold}_{\Acalone}}

\newcommand{\PVWcompAone}{\mathbf{P}_{\hatV_{\Acalone}, \hatW_{\Acalone}}^{\perp}}

\newcommand{\PVcompAone}{\mathbf{P}_{\hatV_{\Acalone}}^{\perp}}

\newcommand{\Mbold}{\mathbf{M}}

\newcommand{\Scal}{\mathcal{S}}
\DeclareMathOperator{\Span}{Span}
\newcommand{\SpanVcal}{\Span(\Vcal)}

\newcommand{\one}{\mathds{1}}
\newcommand{\R}{\mathbb{R}}

\newcommand{\SEhatVAone}{\widehat{\text{SE}}(\VboldAone)}
\newcommand{\epshatVAone}{\hat{\epsbold}(\VboldAone)}
\newcommand{\deltahatVAone}{\hat{\deltabold}(\VboldAone)}
\newcommand{\Tsf}{\top}
\DeclareMathOperator{\E}{\textsf{E}}

\DeclareMathOperator{\Cov}{Cov}

\begin{document}

\title{{TSCI}: two stage curvature identification for causal inference with invalid instruments}  
  
\author[1]{David Carl~\orcidlink{0000-0003-0615-0133}} 
\author[1]{Corinne Emmenegger~\orcidlink{0000-0003-0353-8888}}
\author[1]{Peter B\"uhlmann~\orcidlink{0000-0002-1782-6015}}
\author[2]{Zijian Guo~\orcidlink{0000-0002-2888-7016}}
\affil[1]{Seminar for Statistics, ETH Z\"urich}
\affil[2]{Department of Statistics, Rutgers University}

\date{}

\setcounter{Maxaffil}{0}
\renewcommand\Affilfont{\itshape\small}
\maketitle

\begin{abstract}
\pkg{TSCI} 
  implements treatment effect estimation from observational data under invalid instruments in the \proglang{R} statistical computing environment.
  Existing instrumental variable approaches rely on arguably strong and untestable identification assumptions, which limits their practical application. 
  \pkg{TSCI} does not require the classical instrumental variable identification conditions and is effective even if  
  all instruments are invalid. 
  \pkg{TSCI} implements a two-stage algorithm. 
  In the first stage, machine learning is used to cope with nonlinearities and interactions in the treatment model. 
  In the second stage, a space to capture the instrument violations is selected in a data-adaptive way. These violations are then projected out to estimate the treatment effect. 
\end{abstract}

\proglang{R}.

\section{Introduction: invalid instruments}

Inferring causal treatment effects from observational studies may suffer from endogeneity due to unmeasured confounders. 
A common remedy is to use instrumental variables (IVs) to isolate the variation in the treatment that is uncorrelated with the unmeasured confounders. 
However, valid inference requires these IVs to satisfy stringent and untestable assumptions, i.e., \ref{assumpt:A2}--\ref{assumpt:A3} below. Conditioning on the based covariates, the IVs 
\begin{enumerate}
    \item[{\crtcrossreflabel{(A1)}[assumpt:A1]}] need to be associated strong enough with the treatment variable,  
    \item[{\crtcrossreflabel{(A2)}[assumpt:A2]}] must not be associated with the unmeasured confounders, and 
    \item[{\crtcrossreflabel{(A3)}[assumpt:A3]}] must not directly affect the outcome variable. 
\end{enumerate}
In practice, \ref{assumpt:A2} and~\ref{assumpt:A3} may be questionable or uncheckable, 
and empirical analyses often rely on external knowledge to verify them, which may be prone to errors. 
Therefore, it is crucial to develop methods that are agnostic to these assumptions because falsely relying on them may introduce substantial bias and invalidate inference. 
We consider IVs that violate~\ref{assumpt:A2} or~\ref{assumpt:A3} 
and call such IVs \emph{invalid}. 
To cope with invalid IVs, the following approaches with software for the \proglang{R} environment for statistical computing~\citep{Rlang} exist. 
\begin{itemize}
    \item Assuming near-orthogonality of the effect of the IVs on the treatment and the direct effect of the IVs on the outcome, for which software is available in the supplementary material of~\citet{Bowden2015}. 
    \item Assuming a treatment model with heteroscedastic errors, for which software is available 
    on the github repository of~\citet{TchetgenTchetgen2021}. 
    \item Selecting valid IVs from a pool of potentially invalid ones, for which software is available 
    in the \proglang{R} packages \pkg{RobustIV}~\citep{RobustIV} implementing 
    robust causal inference~\citep{Guo2018,guo2021causal}, 
    \pkg{controlfunctionIV}~\citep{controlfunctionIV} implementing a control function approach~\citep{Guo2020}, and \pkg{CIIV}~\citep{CIIV} implementing the confidence interval method~\citep{CIIV-ref}. See \citet{koo2023robustiv} for the implementation details of 
    \pkg{RobustIV} and \pkg{controlfunctionIV}.
\end{itemize}
An implementation of IV selection can also be found in the  \proglang{Stata}~\citep{Stata} module \pkg{SIVREG}~\citep{Farbmacher2017}. 
Please also see~\citet{GuoBuehlmann2022} for more references to these approaches. 

The two stage curvature identification (TSCI) approach proposed by~\citet{GuoBuehlmann2022} makes none of the restrictions~\ref{assumpt:A1}--\ref{assumpt:A3},  
and all IVs may be invalid. 
The rather mild key assumption is that violations and the association between the IV and the treatment arise from different functional forms.
That is, coincidental cases  where they are all, for instance, linear, are excluded. Machine learning is used to learn the possibly nonlinear treatment model. Our \proglang{R} package, \pkg{TSCI}, provides software for this method. 

For valid IVs, \proglang{R} packages exist to estimate treatment effects from observational data using machine learning, \pkg{DoubleML}~\citep{Chernozhukov2018, DoubleML2021}, and machine learning with additional regularization, \pkg{dmlalg}~\citep{Emmenegger2021, dmlalg}. However, they use a linear fitting between the IV and the treatment.
In contrast, \pkg{TSCI} 
can cope with invalid instruments, and it uses machine learning to capture complex nonlinearities and interaction terms among the IVs and 
covariates in the treatment model. 

This paper is organized as follows. Section~\ref{sect:TSCI} presents the statistical model and discusses key steps of the \pkg{TSCI} algorithm on a theory level. Section~\ref{sect:using-tsci} explains how to use \pkg{TSCI} to estimate and make inference for the treatment effect from observational data under invalid IVs using empirical data. 
Section~\ref{sect:guide} summarizes \pkg{TSCI}'s main functionality, and Section~\ref{sect:summary} finally concludes.

\section{Two stage curvature identification}\label{sect:TSCI}

Subsequently, we present the TSCI methodology of~\citet{GuoBuehlmann2022}. 
We have $i=1,\ldots,n$ independent and identically distributed observations as follows. 
For IVs $\Zi$ and exogenous baseline covariates
$\Xi$, 
we consider the treatment model
\begin{equation}\label{eq:eqTRT}
    \Di=f(\Zi,\Xi)+ \deltai, \quad \E[\deltai | \Zi, \Xi] = 0
\end{equation}
for some unknown function $f$, and we consider the outcome model
\begin{displaymath}
    \Yi=\beta\Di + g(\Zi,\Xi) + \epsi, \quad \E[\epsi | \Xi,\Zi] = 0
\end{displaymath}
for some unknown function $g$. 
The error terms $\deltai$ and $\epsi$ may be correlated, which introduces endogeneity.
If the IVs $\Zi$ were valid, the function $g$ would not directly depend on $\Zi$.
We aim to estimate and make inference for the effect $\beta$ of the treatment on the outcome. 
The variables $\Xi$ and $\Zi$ may be multi-dimensional. All other variables, and thus also $\beta$, are \mbox{$1$-dimensional}. 
Introducing $\phi(\Xi) = \E[g(\Zi,\Xi)|\Xi]$, we can rewrite the outcome model as
\begin{displaymath} 
    \Yi=\beta\Di + h(\Zi,\Xi) + \phi(\Xi) + \epsi, \quad \E[\epsi | \Xi,\Zi] = 0 
\end{displaymath}
with $h(\Zi,\Xi) = g(\Zi,\Xi)-\E[g(\Zi,\Xi)|\Xi]$. 
If the IVs $\Zi$ were valid, we had $h\equiv 0$ because $g$ would not directly depend on $\Zi$. Conversely, if $h$ is non-zero, 
then~\ref{assumpt:A2} or~\ref{assumpt:A3} are violated.
Consequently, $h$ is a measure of IV violation, and we call it \emph{violation function}. 

The crucial idea of TSCI to estimate $\beta$ is to make use of differences in the functional forms of the treatment model 
$f$ and the violation function $h$.
Assume that $f$ can be approximated well by some subspace $\Scal$ of some Hilbert space $S$ and that $g = h + \phi$ can be approximated well by some subspace $\SpanVcal \subseteq S$ spanned by some set of basis functions $\Vcal$. 
To cope with the IV violation, we need to project out from the outcome model the part of $g$ that is ``non-orthogonal'' to $f$. That is, it is not necessary to project out all of $g$, thus all of $\SpanVcal$, but it suffices to project to a space where $f$ and $g$ become ``orthogonal''. That is, we choose some space $\VOmega$ satisfying $\Scal \cap \SpanVcal \subseteq \VOmega \subseteq S$, and we will project to the orthogonal complement of $\VOmega$. We call $\VOmega$ \emph{violation space}, and an example of such a $\VOmega$ is $\SpanVcal$ itself. Because $\VOmega$ is unknown in practice, \pkg{TSCI} selects one out of user-specified candidates in a data-adaptive way.
Provided enough variation in the approximation of $f$ using $\Scal$ remains after projecting it onto the orthogonal complement of $\VOmega$, we can use these projected values to overcome the IV violation. In particular, if $h$ is linear in $\Zi$ and $f$ is quadratic in $\Zi$, we could use the nonlinear part of $f$ in $\Zi$ to estimate $\beta$. Note that in~\citet{GuoBuehlmann2022}, $\VOmega$ is not explicitly defined but is chosen as $\SpanVcal$. 

To estimate the treatment effect $\beta$, we require approximating two objects: the function $f$ and the violation space $\VOmega$, which may depend on the method chosen to approximate $f$.
The function $f$ is estimated using machine learning (or a basis expansion), $\VOmega$ is selected by testing user-specified candidates for which the resulting (generalized) IV strength is large enough. In particular, the candidates differ in the basis used to approximate $h$ but contain the same basis to approximate $\phi$. The TSCI algorithm consists of the following two stages:
\begin{itemize}
    \item Using machine learning to predict the treatment in the treatment model~\eqref{eq:eqTRT}. 
    \item Rescaling the outcome by a ``hat matrix'' that comes from the above prediction step, projecting the rescaled response and predicted treatment to the orthogonal complement of the violation space (where IV violation is no longer a problem), and performing ordinary least squares. 
\end{itemize}
An additional data splitting strategy is required if machine learning is used to approximate $f$. 
In the classic two stage least squares (TSLS) algorithm~\citep{Theil1953a, Theil1953b}, the responses need no rescaling in the second step. \pkg{TSCI} requires it because the ``hat matrix'', in contrast to TSLS, may not be orthogonal. 
The next sections detail sample splitting, the basis expansion of $\phi$, and the estimation of  $\SpanVcal$, $f$, and finally $\beta$. Subsequently, we denote in bold row-wise concatenations of the observations; for instance, $\hbold$ represents the vector $(h(Z_1,X_1),\ldots,h(Z_n,X_n))^{\Tsf}$.

\subsection{Sample splitting and machine learning estimator of $f$}\label{sect:estimating-f}

If a flexible machine learning algorithm such as random forest or boosting is used to fit the treatment model, sample splitting is essential to remove bias due to overfitting. Sample splitting prevents overfitting; because in the case of overfitting, endogeneity of the treatment may not be fully removed. 
We partition the sample indices into two sets $\Acalone$ and $\Acaltwo$. By default, they are approximately of size $2n/3$ and $n/3$, respectively, with $\Acalone = \{1,2,\ldots, 2n/3\}$ without loss of generality. 
An estimator of $f(\Zi,\Xi)$ for $i\in\Acalone$ is constructed by applying some machine learning algorithm on the data $\Acaltwo$ and by then predicting it on $\Acalone$. We call these predictions $\hatfAonebold$. 
That is, the treatments $\DAtwobold$ are not used in this step, where $\DAtwobold$ contains the entries of $\D$ that are in $\Acaltwo$. 
The estimated $\hatfAonebold$ must be of the form ``hat matrix'' times treatment vector, denoted by $\hatfAonebold = \Omegabold\DAonebold$ for some not necessarily orthogonal matrix $\Omegabold\in\R^{|\Acalone|\times|\Acalone|}$. 
Such a representation is feasible for random forests if we interpret the forest as a weighted nearest neighbor method and use the induced weighting function for prediction~\citep{Lin-Jeon2006}. 
Indeed, for a random forest with $S$ decision trees, the entry $(i,j)$ of $\Omegabold$ for $i,j\in\Acalone$ 
is given by
\begin{displaymath}
    \Omegabold_{i,j} 
    = \frac{1}{S}\sum_{s=1}^S
    \omega_j(\Zi,\Xi,\theta_s)
    \quad\mathrm{for\ weights}\quad
    \omega_j(z,x,\theta_s) = \frac{\one_{\{(\Zj,\Xj)\in\Rcal_{l(z,x,\theta_s)}\}}}{\sum_{k\in\Acalone} \one_{\{(\Zk,\Xk)\in\Rcal_{l(z,x,\theta_s)}\}}},
\end{displaymath}
where $\Rcal_{l(z,x,\theta_s)}$ denotes the leaf of the $s$th tree that contains $(z,x)$, and $\theta_s$ denotes a random parameter that determines how the $s$th tree was grown. 
A similar representation also holds for $L_2$ boosting~\citep{Buhlmann-Hothorn2017, Buhlmann-Yu2006} 
with regression trees as base learners. 

In \pkg{TSCI}, three methods are implemented to compute $\Omegabold$: random forest, $L_2$ boosting with regression trees as base learners, and polynomial basis expansion. 
The former two approaches are flexible 
and might capture complex functions $f$ containing nonlinearities and interactions. This is desirable because
if more variation in $f$ is captured, which machine learning algorithms usually achieve, then the resulting IV strength 
is larger. The \emph{(generalized) IV strength} measures the ratio of the variation of the treatment and the estimated treatment error variance in $\Acalone$ after adjusting for the violation space candidate~\citep{GuoBuehlmann2022}.
If more variation in $f$ that is orthogonal to the violation is captured, 
enough variation in $f$ 
remains after projecting out the violation space. 
Preserving variation improves the efficiency of the method and gives us more room to test for IV invalidity. 
When testing for invalidity, our goal is to select the smallest possible violation space that can address the violation. Testing for larger spaces requires a higher IV-strength. 
Using a polynomial approach to approximate $f$ can be helpful in the special situation when the part of $f$ that is orthogonal to the violation can be captured well by polynomials. 
In this case, the violation space sequence selection becomes particularly simple because one only has to consider subspaces spanned by elements of the respective polynomial basis. However, a polynomial expansion cannot be employed for binary IVs. 

Apart from the three approaches to estimate $\Omegabold$ that are implemented in \pkg{TSCI}, the user may provide an individual matrix. 
Furthermore, sample splitting is only necessary if a machine learning method is used to estimate $f$ and is not required for the polynomial basis approach. Below, we use sample splitting notation.

\subsection{Violation space selection and basis expansion}\label{sect:basis-violation}

We use a set of basis function $\basish$ to approximate the violation function $h$ and another set of basis functions $\basisphi$ to approximate $\phi$ (more precisely, the part thereof that is not orthogonal to $f$, that is, $\Scal\cap \SpanVcal$) on $\Acalone$.

We choose a violation space in a data-dependent way. Given a sequence of $Q+1$ many ideally nested sets of basis functions $\{\Vcal_q\}_{0\le q\le Q}$ such that $\Vcal_q \subset (\basish \cup \basisphi)$ for every ${0\le q\le Q}$ and $\Vcal_0 = \basisphi$, we estimate the treatment effect $\beta$ for each candidate $\Vcal_q$, $0\leq q \le  \Qmax \le Q$, where $\Qmax$ denotes the last candidate for which all $q \le  \Qmax$ candidates were considered to provide enough IV strength. 
Subsequently, each treatment effect estimator
is tested against all its successors for significant differences. 
Finally, the first violation space candidate $\Vcal_{\qcomp}$ for which no significant differences can be found compared to all $\Vcal_{q'}$, $\qcomp < q' \leq \Qmax$ 
is selected, where the subscript ``comp'' stands for comparison. 
That is, the smallest space is selected from where onward the hypothesis of having obtained an unbiased treatment estimate could not be rejected.
This approach also tests if~\ref{assumpt:A2} or~\ref{assumpt:A3} are violated at all because the first candidate, $\Vcal_0 = \basisphi$, represents having no violation of~\ref{assumpt:A2} 
or~\ref{assumpt:A3}. 
Consequently, if $\Vcal_0$ is not selected, violations are present because the hypothesis of having no violation was rejected.
A bootstrap test is used to test if the IV strength exceeds some threshold and are consequently strong enough~\citep{GuoBuehlmann2022}. This test can also be used in case of heteroscedastic errors. Its explicit formulation is given in~\citet{GuoBuehlmann2022}.

In finite samples, some violations might not be detected by this approach.
This may happen if the remaining bias due to the violation after adjusting for the selected violation space 
is small enough such that the treatment 
estimates are not found to be significantly different. 
Therefore, a more conservative approach to select a suitable violation space is also implemented in \pkg{TSCI}. We refer to the former approach described above as the comparison method and the latter as the conservative method. 
If the IVs are strong enough, the conservative method does not select the first violation space candidate for which no significant differences were found, but its successor in the sequence, which makes the method more conservative. 

We leave the specification of a suitable basis to approximate $\phi$ to the user. In case the user does not have good knowledge about a suitable basis, 
they can use the polynomial basis expansion approach implemented in \code{TSCI\_poly}, omit the covariates, or treat them as additional IVs that might be invalid.

\subsection{Estimating the treatment effect}\label{sect:estimating-beta}

Finally, we concatenate the basis functions to approximate  $\hbold_{\Acalone}$ and the basis functions to approximate $\phibold_{\Acalone}$ row-wise into the \emph{violation matrix} $\Vbold_{\Acalone}$.
Consequently, $\VboldAone$ is understood to contain a basis of $\Scal \cap \SpanVcal$ and we denote the $\WboldAone$ as the submatrix of $\VboldAone$ containing the basis to approximate $\phibold_{\Acalone}$.
The treatment effect estimator is then given by projecting $\Omegabold\Y_{\Acalone}$ and $\hatfAonebold=\Omegabold\D_{\Acalone}$ onto the orthogonal complement of the space spanned by $\hatVAone=\Omegabold\VboldAone$, and by afterward performing ordinary least squares with the projected quantities. 
The rescaling of the violation matrix $\VboldAone$ with $\Omegabold$ is due to rescaling $\YAonebold$ with $\Omegabold$, and the latter is required because $\Omegabold$ is not necessarily orthogonal. 
The space spanned by $\VboldAone$ is an approximation of $\VOmega$.  
Additionally, a bias correction term, which is explicitly given in~\citet{GuoBuehlmann2022}, is subtracted from this effect size estimator to correct for any remaining correlation of $\epsboldAone$ and $\DAonebold$ due to endogeneity, which results in the final estimator
\begin{equation}\label{eq:beta-estimator}
    \betahat = 
    \frac{\YAonebold^{\Tsf} \Omegabold^{\Tsf} \PVcompAone \Omegabold \DAonebold}
    {\DAonebold^{\Tsf} \Omegabold^{\Tsf} \PVcompAone \Omegabold \DAonebold} -
    (\mathrm{bias\ term}).
\end{equation}
To motivate this estimator, we rewrite its first term in~\eqref{eq:beta-estimator} as
\begin{equation}\label{eq:beta-expansion}
\begin{aligned}
    \beta \cdot \frac{\DAonebold^{\Tsf} \Omegabold^{\Tsf} \PVcompAone \Omegabold \DAonebold}
    {\DAonebold^{\Tsf} \Omegabold^{\Tsf} \PVcompAone \Omegabold \DAonebold} + 
    \frac{\hboldAone^{\Tsf} \Omegabold^{\Tsf} \PVcompAone \Omegabold \DAonebold}
    {\DAonebold^{\Tsf} \Omegabold^{\Tsf} \PVcompAone \Omegabold \DAonebold} + \\
    \frac{\phiboldAone^{\Tsf} \Omegabold^{\Tsf} \PVcompAone \Omegabold \DAonebold}
    {\DAonebold^{\Tsf} \Omegabold^{\Tsf} \PVcompAone \Omegabold \DAonebold} +
    \frac{\epsboldAone^{\Tsf} \Omegabold^{\Tsf} \PVcompAone \Omegabold \DAonebold}
    {\DAonebold^{\Tsf} \Omegabold^{\Tsf} \PVcompAone \Omegabold \DAonebold}.
\end{aligned}
\end{equation}
The first term in~\eqref{eq:beta-expansion}  simplifies to the true causal effect $\beta$. 
The second term in~\eqref{eq:beta-expansion} is the bias due to having invalid IVs, but it is small if the part of $\hboldAone$ that is not orthogonal to $\hatfAonebold$ is approximated well by the column space of the violation matrix $\VboldAone$. 
Analogously, the third term in~\eqref{eq:beta-expansion} is small if the part of $\phiboldAone$ that is not orthogonal to $\hatfAonebold$ is approximated well by the column space of $\VboldAone$ or more specifically by the column space of the submatrix $\WboldAone$.
Lastly, the bias correction term in~\eqref{eq:beta-estimator} corrects for an potential 
overfitting bias caused by the fourth term in~\eqref{eq:beta-expansion} because $\epsboldAone$ and $\DAonebold$ are correlated if endogenous confounding is present.  
In the case of overfitting, part of this correlation might remain even after rescaling $\epsboldAone$ and $\DAonebold$ by $\Omegabold$ and subsequently projecting onto the orthogonal complement of $\hatVAone$.

\subsection{Inference}\label{sect:inference}

If $\VboldAone$ is approximating the parts of $\hboldAone$ and $\phiboldAone$ that are not orthogonal to $\hatfAonebold$ well and the IVs are strong enough, it can be shown that $( \betahat - \beta ) / \SEhatVAone  \xrightarrow{d} \mathcal{N}(0, 1)$ under some regularity conditions, 
where $\SEhatVAone$ denotes an estimator of the standard deviation of the treatment effect estimator. The user can choose between using the estimator of the standard deviation proposed in \citet{GuoBuehlmann2022} or an alternative bootstrap approach (see appendix \ref{sect:appendix-b}). The bootstrap approach is motivated by its potential to provide more accurate finite-sample tests and confidence intervals, in line with classical theory and based on some empirical evidence. In either way, the asymptotic Gaussian approximation of the treatment effect estimator is used to compute confidence intervals and $p$~values.

\subsection{Counteracting the randomness of sample splitting}\label{sect:repeated-splitting}

Splitting the sample in two parts is random and may affect the treatment effect estimator. Thus, the whole procedure is repeated multiple times. The treatment effect estimates from the different repetitions are aggregated by the median, and the user may choose between two schemes to compute aggregated $p$~values and confidence intervals (and standard errors in the ``DML'' scheme presented below).
The first method, called ``FWER'' in \pkg{TSCI}, leads to valid $p$~values by controlling family-wise error rates and has initially been proposed by~\citet{Meinshausen2012}. The obtained $p$~values may be conservative. 
The second method, called ``DML'' in \pkg{TSCI}, directly aggregates the standard error estimators from the individual repetitions and has initially been proposed by~\citet{Chernozhukov2018}.

\section{Using TSCI} \label{sect:using-tsci}

Subsequently, we illustrate the use of \pkg{TSCI} using two empirical datasets.

\subsection{Example 1} \label{sect:using-tsci-ex1}

The first step in applying \pkg{TSCI} is to select a learner for the treatment model.
\pkg{TSCI} offers four built-in options, and they are given in Table~\ref{tab:overview-TSCI}.
Their most important input arguments are presented in Table~\ref{tab:arguments-TSCI}. 
\begin{table}[ht]
\renewcommand{\arraystretch}{1.25}
\centering
\begin{tabular}{lp{11.4cm}}
\hline
Function                    & Description \\ \hline
\code{tsci\_forest}          & Uses random forests to fit the treatment model.
                            Requires the user to specify a set of violation space candidates.\\
\code{tsci\_boosting}        & Uses boosting to fit the treatment model. 
                            Requires the user to specify a set of violation space candidates.\\
\code{tsci\_poly}            & Uses polynomial basis expansion 
to fit the treatment model.
                            Does not require the user to specify a set of violation space candidates. \\
\code{tsci\_secondstage}      & Does not fit the treatment model but uses a user-provided hat matrix instead.
                            Requires the user to specify a set of violation space candidates. \\ \hline
\end{tabular}
\caption{\label{tab:overview-TSCI} \pkg{TSCI}'s four main functions for treatment effect size estimation.}
\end{table}

\begin{table}[h!]
\renewcommand{\arraystretch}{1.25}
\centering
\begin{tabular}{lp{10.2cm}}
\hline
Arguments                    & Description \\ \hline
\code{Y}                     & Numeric vector of outcomes.\\
\code{D}                     & Vector of treatments.\\
\code{Z}                     & Matrix or data frame of instruments.\\
\code{X}                     & Matrix or data frame of covariates for the treatment model. 
\\
\code{W}                     & (Transformed) observations of baseline covariates \code{X} for the outcome model. \\
\hline
\code{vio\_space}             & List to specify the violation space candidates. \\
\code{create\_nested\_sequence}& Logical. If \code{TRUE}, the violation space candidates are defined sequentially starting with \code{W} and subsequently adding the next element of \code{vio{\_}space}. If \code{FALSE}, the violation space candidates are defined as by adding \code{W} to the elements of \code{vio\_space}. \\
\code{sel\_method}            & Selection method to estimate the treatment effect. Either \code{"comparison"} or \code{"conservative"}; see Section~\ref{sect:basis-violation}.\\
\code{split\_prop} & Proportion of observations used to fit the outcome model. Has to be a value in $(0, 1)$.\\
\code{sd\_boot}              & Logical. if \code{TRUE}, it determines the standard error using a bootstrap approach. \\
\code{iv\_threshold}          & Minimal value of threshold of IV strength test.\\
\code{threshold\_boot}        & Logical. If \code{TRUE}, the threshold of the IV strength is determined using a bootstrap approach to adjust for estimation error of the IV strength, which leads to a more conservative IV strength test. 
\\
\hline
\code{nsplits}               & Number of data splits.\\
\code{mult\_split\_method}     & Method to calculate  standard errors, $p$~values, and to construct the confidence intervals if multi-splitting is performed. Either \code{"FWER"} or \code{"DML"}; see Section~\ref{sect:repeated-splitting}.\\
 \hline
\end{tabular}
\caption{\label{tab:arguments-TSCI} 
Most important arguments of \pkg{TSCI}'s four main functions in Table~\ref{tab:overview-TSCI}.}
\end{table}

Subsequently, we demonstrate the use of \pkg{TSCI} using the dataset from~\citet{Card1993}. This dataset consists of $n=3010$ observations, and the aim is to estimate the causal effect of education (\code{educ}) on log earnings (\code{lwage}) using proximity to a 4-year college (\code{nearc4}) as an instrument. 
As exogenous baseline covariates \code{X}, we consider  experience (\code{exper}, \code{expersq}), race (\code{race}), and geographical information (\code{smsa}, \code{smsa1966}, \code{south}, \code{reg661}--\code{reg668}).
The dataset is available in the \proglang{R}-package \pkg{ivmodel}~\citep{ivmodel} and has also been analyzed by~\citet{GuoBuehlmann2022}. 
\begin{Schunk}
\begin{Sinput}
R> library("TSCI")
R> library("ivmodel")
R> data("card.data")
\end{Sinput}
\end{Schunk}
The first step in applying \pkg{TSCI} is to choose one of the functions from Table~\ref{tab:overview-TSCI}. Below, we demonstrate how to use \code{tsci\_forest} and afterwards \code{tsci\_secondstage}. The functions \code{tsci\_boosting} and \code{tsci\_poly} are structured analogously to \code{tsci\_forest}. 
The second step in applying \pkg{TSCI} is to select a suitable transformation of the baseline covariates to approximate $\phi$. The matrix of transformed covariates can be passed to the input argument \code{W}. This matrix will be part of every violation space candidate. If \code{W} is not specified, the untransformed baseline covariates are added instead. Finally, a list containing different candidates to approximate $h$ must be provided. By default, \code{tsci\_forest} creates a nested sequence of violation space candidates from this list by starting with \code{W} as the first violation space candidate and subsequently adding the next element of the provided list to the current violation matrix to create the next violation space candidate.
For instance, to specify polynomials up to degree three in the instrument \code{Z} together with the untransformed baseline covariates \code{X} as the violation space candidates, it suffices to pass a list with the elements
\begin{math} \{ $\code{Z}$,\:  $\code{Z}$^2,\: $\code{Z}$^3 \} \end{math} to \code{tsci\_forest}. This notation holds irrespective of the number of columns of \code{Z}. 
Alternatively, the function \code{create\_monomials}  automatically creates this sequence by calling \code{create\_monomials(Z = Z, degree = 3)}. 
In our demonstration of~\pkg{TSCI}, we are going to consider a list of length two where the first list element corresponds to the instrument \code{nearc4} and the second element to the interactions of \code{nearc4} with the baseline covariates:
\begin{Schunk}
\begin{Sinput}
R> vio_space <- with(card.data, list(nearc4,
+    nearc4 * cbind(exper, expersq, black, south, smsa, smsa66, reg661, 
+    reg662, reg663, reg664, reg665, reg666, reg667, reg668)))
R> ## alternatively: 
R> # vio_space <- create_interactions(card.data$nearc4, as.matrix(X))
\end{Sinput}
\end{Schunk}
There is another function \code{create\_interactions} that can be used to create the same list; please see the comment in the above \proglang{R} code.
To use non-nested violation space candidates, we would have to set the input parameter
\code{create\_nested\_sequence} of \code{tsci\_forest} to \code{FALSE}, in which case each element of the provided list will be treated
as a violation space candidate itself without adding the previous ones. After specifying the violation space candidates, we can call
\code{tsci\_forest} to estimate the treatment effect:
\begin{Schunk}
\begin{Sinput}
R> set.seed(10)
R> Xname <- c("exper", "expersq", "black", "south", "smsa", "smsa66", 
+    "reg661", "reg662", "reg663", "reg664", "reg665", "reg666", 
+    "reg667","reg668")
R> fit_forest <- tsci_forest(Y = card.data$lwage, D = card.data$educ,
+    Z = card.data$nearc4, X = card.data[, Xname], vio_space = vio_space)
\end{Sinput}
\end{Schunk}
%
There is a \code{summary} method to provide an overview of the most relevant statistics, and it
has the same form for \code{tsci\_forest}, \code{tsci\_boosting}, \code{tsci\_poly}, and
\code{tsci\_secondstage}. 
\begin{Schunk}
\begin{Sinput}
R> summary(fit_forest)
\end{Sinput}
\begin{Soutput}
Statistics about the data splitting procedure:
Sample size A1: 2007 
Sample size A2: 1003 
Number of data splits: 500 
Aggregation method: FWER 

Statistics about the validity of the instrument(s):
       valid      invalid non_testable 
         262          238            0 

Treatment effect estimate of selected violation space candidate(s):
              Estimate Std_Error    2.5%   97.5% Pr(>|t|)
TSCI-Estimate  0.05752         . 0.02703 0.08757    4e-05
Selection method: comparison 

Statistics about the treatment model:
Estimation method: Random Forest 

Statistics about the violation space selection:
   q_comp q_cons Qmax
q0    262      0    0
q1     41    262    0
q2    197    238  500
\end{Soutput}
\end{Schunk}

The first part of the above summary output provides information about the data splitting procedure. The number of data points to estimate the treatment model (\code{A1}) and the outcome model (\code{A2}) are given first. Their size can be adjusted via the input argument \code{split\_prop}. 
If not specified otherwise,  $10$ data splits (input argument \code{nsplits}) are performed to counteract the randomness introduced by the data splitting, and the \code{"FWER"} method (input argument \code{mult\_split\_method}) is used to aggregate the estimators obtained from each data split. 
Next, the result of testing for invalidity of the IV is presented. 
In the above output, in $262$  of the $500$ data splits  performed by \code{tsci\_forest}, the IVs  
 were found to be valid, and $238$ were invalid. That is, assumptions~\ref{assumpt:A2} or~\ref{assumpt:A3} were found to be violated in $238$ cases.
The number of data splits for which the IV strength was too weak to estimate the treatment effect
for all violation space candidates apart from the space assuming no violation are listed under \code{non\_testable}. 
In our example, this did not occur in any of the $500$ data splits. Next, the treatment effect
obtained by the selected violation space candidate(s) and the method used to obtain the hat matrix $\Omegabold$ are displayed. By default, the bootstrap approach is used to obtain the standard errors.
Finally, the number of times (out of the $500$) each violation space candidate was selected by the comparison method (first column) and the conservative method (second column) are displayed alongside with the number of times each violation space candidate was found to be the largest violation space candidate for which the IVs were strong enough (third column). 
In cases where this violation space candidate coincides with the violation space candidate selected by the comparison method (and consequently also by the conservative method), the treatment effect estimate should be interpreted carefully. Indeed, 
the bias caused by a violation of  
assumptions~\ref{assumpt:A2} or~\ref{assumpt:A3} 
might not be fully addressed because it is not possible to reliably test whether the treatment effect estimate would change significantly when using a larger violation space candidate.
The user can choose whether the treatment effect should be estimated using the comparison method or the conservative method by specifying the input parameter \code{sel\_method}. By default, the comparison method is used, as was the case for this example. 
To obtain a more detailed summary of the output, we can set  \code{extended\_output = TRUE}.
Then, the treatment effect estimate, the estimated IV strength, and the threshold for considering IVs as being strong enough are specified for each violation space candidate. If more than one data split is performed, the treatment effect estimate obtained by the selected violation space candidates might not coincide with any of those point estimates as in each data split a different violation space candidate might be selected. We can see that this is indeed the case here.
\begin{Schunk}
\begin{Sinput}
R> summary(fit_forest, extended_output = TRUE)
\end{Sinput}
\begin{Soutput}
Statistics about the data splitting procedure:
Sample size A1: 2007 
Sample size A2: 1003 
Number of data splits: 500 
Aggregation method: FWER 

Statistics about the validity of the instrument(s):
       valid      invalid non_testable 
         262          238            0 

Treatment effect estimate of selected violation space candidate(s):
              Estimate Std_Error    2.5%   97.5% Pr(>|t|)
TSCI-Estimate  0.05752         . 0.02703 0.08757    4e-05
Selection method: comparison 

Treatment effect estimates of all violation space candidates:
        Estimate Std_Error    2.5%   97.5% Pr(>|t|)
TSCI-q0  0.06293         . 0.03389 0.09196  0.00000
TSCI-q1  0.06185         . 0.03243 0.09179  0.00001
TSCI-q2  0.05102         . 0.01925 0.08296  0.00077

Statistics about the treatment model:
Estimation method: Random Forest 

Statistics about the violation space selection:
   q_comp q_cons Qmax
q0    262      0    0
q1     41    262    0
q2    197    238  500

Statistics about the IV strength:
   IV_Strength IV_Threshold
q0      130.08           40
q1      126.78           40
q2       99.98           40
\end{Soutput}
\end{Schunk}
Other methods to extract relevant information are \code{coef}, which returns the treatment effect estimate, and \code{cofint}, which returns the confidence interval of the treatment effect estimate at the specified confidence level.

If we have reason to believe that the untransformed baseline covariates are not able to approximate $\phi$ sufficiently well, 
we can set the function argument \code{W} of \code{tsci\_forest}. In that case, the original baseline covariates \code{X} are used to fit the treatment model and the transformed baseline covariates \code{W} are used for the outcome model.
For instance, column-wise concatenations of basis splines for the individual covariates may be considered:  
\begin{Schunk}
\begin{Sinput}
R> suppressPackageStartupMessages(library("fda"))
R> X <- card.data[, c("exper", "expersq")]
R> head(X, 4)   # covariates
\end{Sinput}
\begin{Soutput}
  exper expersq
1    16     256
2     9      81
3    16     256
4    10     100
\end{Soutput}
\begin{Sinput}
R> nknots <- 2  # number of knots for bspline basis
R> norder <- 3  # order of spline
R> nbasis <- nknots + norder - 2 # number of basis functions
R> W <- matrix(ncol = NCOL(X) * nbasis, nrow = NROW(X))
R> for (j in seq_len(NCOL(X))) {
+    knots <- quantile(unique(X[, j]), seq(0, 1, length = nknots))
+    basis <- create.bspline.basis(rangeval = range(knots), breaks = knots,
+      norder = norder)
+    W[, c(((j - 1) * nbasis + 1) : (j * nbasis))] <- eval.basis(X[, j], basis)
+  }  
R> head(W, 4) # basis transformation of covariates
\end{Sinput}
\begin{Soutput}
          [,1]      [,2]      [,3]      [,4]      [,5]       [,6]
[1,] 0.0926276 0.4234405 0.4839319 0.2663262 0.4994836 0.23419013
[2,] 0.3705104 0.4763705 0.1531191 0.7172073 0.2593473 0.02344546
[3,] 0.0926276 0.4234405 0.4839319 0.2663262 0.4994836 0.23419013
[4,] 0.3194707 0.4914934 0.1890359 0.6576627 0.3066027 0.03573458
\end{Soutput}
\end{Schunk}
%

\subsubsection{Specifying an individual hat matrix}

The function \code{tsci\_secondstage} offers the flexibility to specify an individual hat matrix $\Omegabold$. 
This can be advantageous if we have good knowledge about the treatment model and can thus specify a model that captures more treatment variation than a random forest or boosting approach would, 
or if we want to project the violation into a certain space (for example, using an additive model to avoid the need to test for
interactions in the violation).

We illustrate the usage of \code{tsci\_secondstage} using again the dataset from~\citet{Card1993}.
\citet{Card1993} discuss the possibility that their initially chosen instrument, college proximity, varies by family background. In such a case, college proximity would not be a valid instrument because it has an effect on earnings apart from the effect through education.  
To illustrate the use of \code{tsci\_secondstage}, we are going to explore if the interaction between college proximity and family background is a valid instrument. To do this, we specify the treatment model as a linear combination of the baseline covariates and the new instrument.
Then, we can use \code{tsci\_secondstage} to test for a violation of assumptions~\ref{assumpt:A2} or~\ref{assumpt:A3} 
and to obtain a treatment effect estimator. 
Following~\citet{Card1993}, we use the predicted education level in absence of a nearby college based on race, geographic information in 1966, and family composition information at age 14 as our
proxy for family background. We refer to~\citet{Card1993} for the details how to 
obtain this predicted education level. 
In a first step, we specify the hat matrix $\Omegabold$ and violation space candidates. Here, we use the projection matrix from an OLS fit resulting from regressing education on the interaction instrument and the covariates \code{X} including an intercept.
We specify the violation space candidates to test for a violation of assumptions~\ref{assumpt:A2} or~\ref{assumpt:A3} 
of proximity to college. Because we did not include interactions in the treatment model specification, we also do not need to include them as violation space candidates. 
\begin{Schunk}
\begin{Sinput}
R> Z <- with(card.data,
+            cbind(nearc4,
+                  nearc4 * family_background))
R> vio_space <- list(card.data$nearc4)
R> Xname <- c("exper", "expersq", "black", "south", "smsa", "smsa66",
+    "reg661","reg662", "reg663", "reg664", "reg665", "reg666", "reg667", 
+    "reg668", "fatheduc", "fatheduc_na","motheduc", "motheduc_na", 
+    "parenteduc", "momdad14", "sinmom14", "step14")
R> X <- card.data[, Xname]
R> X$motheduc_na <- as.numeric(X$motheduc_na)
R> X$fatheduc_na <- as.numeric(X$fatheduc_na)
R> A <- cbind(1, as.matrix(Z), as.matrix(X))
R> omega <- A %*% chol2inv(chol(t(A) %*% A)) %*% t(A)
\end{Sinput}
\end{Schunk}
Next, we call \code{tsci\_secondstage}, passing our hat matrix to the input parameter \code{weight}.
\begin{Schunk}
\begin{Sinput}
R> fit_secondstage <- tsci_secondstage(Y = card.data$lwage, 
+    D = card.data$educ, Z = Z, W = X, 
+    vio_space = vio_space, weight = omega)
R> summary(fit_secondstage, extended_output = TRUE)
\end{Sinput}
\begin{Soutput}
Statistics about the data splitting procedure:
Sample size: 3010 
No sample splitting was performed.

Statistics about the validity of the instrument(s):
       valid      invalid non_testable 
           0            0            1 

Treatment effect estimate of selected violation space candidate(s):
              Estimate Std_Error    2.5%   97.5% Pr(>|t|)
TSCI-Estimate  0.13125   0.03452 0.06359 0.19891  0.00014
Selection method: comparison 

Treatment effect estimates of all violation space candidates:
        Estimate Std_Error    2.5%   97.5% Pr(>|t|)
TSCI-q0  0.13125   0.03452 0.06359 0.19891  0.00014
TSCI-q1  0.12508   0.04525 0.03639 0.21377  0.00571

Statistics about the treatment model:
Estimation method: Specified by User 

Statistics about the violation space selection:
   q_comp q_cons Qmax
q0      1      1    1
q1      0      0    0

Statistics about the IV strength:
   IV_Strength IV_Threshold
q0       40.21        38.30
q1       25.24        34.93
\end{Soutput}
\end{Schunk}
As shown in the last part of the summary output, the instrumental strength was not large enough to reliably test for a violation of assumptions~\ref{assumpt:A2} or~\ref{assumpt:A3}
using this treatment model specification.

\subsection{Example 2} \label{sect:using-tsci-ex2}

Subsequently, we demonstrate the use of \pkg{TSCI} to 
assess the validity of instruments in a second empirical dataset. 
\citet{Angrist1991} 
estimate the effect of compulsory schooling on log weekly earnings in the United States of men 
born between 1920 and 1929~\citep{AngristData}. 
Compulsory schooling laws require students to remain in school until they reach the legal dropout age. Consequently, students born at the beginning of the year, who start school at an older age, may drop out at an earlier point in their educational career. 
\citet{Angrist1991} estimate the effect of education variation due to differences in quarter of birth, which is due to compulsory schooling laws, on earnings. 
Based on data from college students, \citet{Angrist1991} infer that the quarter of birth does not affect earnings for reasons other than compulsory schooling. 
However, 
assumptions~\ref{assumpt:A2} or~\ref{assumpt:A3} 
may be violated because the instrument may be associated with unmeasured confounders that are associated with both education and earnings such as ability and other factors. Therefore, our aim is to evaluate with \pkg{TSCI} whether this instrument is actually valid or not. To do this, we consider a violation that is linear in the instruments.

We consider the following variables. The response \code{Y} is the log weekly wage (\code{LWKLYWGE}), 
the treatment variable \code{D} the number of years of schooling (\code{EDUC}), and the instrument consists of $30$ dummies indicating an interaction between $3$ birth quartals and $10$ birth years.
\citet{Angrist1991} consider this instrument for education to incorporate cross-year seasonal variation in education.
Their estimate also includes ten year dummies in the wage equation so that the variability in education used to identify the return to education in their TSLS estimates is solely due to differences by season of birth. 
The observed baseline covariates \code{X} consist of $9$ dummy columns indicating the birth year, 
the age (\code{AGE}), 
the marital status (\code{MARRIED}; $1$ corresponds to being married)
a binary indicator for race (\code{RACE}; $1$ corresponds to black), 
an indicator of working in an SMSA (\code{SMSA}; $1$ corresponds to city center), and
eight region of residence dummies (\code{NEWENG}, \code{MIDATL}, \code{ENOCENT}, \code{WNOCENT}, \code{SOATL}, \code{ESOCENT}, \code{WSOCENT}, \code{MT}).
We loaded the complete data~\citep{AngristData} as the data frame \code{df} into our session, and the variables we consider are the following ones:

\begin{Schunk}
\begin{Sinput}
R> # 3 quarter of birth dummies
R> QOB <- as.matrix(df[, c("QTR2", "QTR3", "QTR4")])
R> # 10 year of birth dummies
R> YOB <- as.matrix(df[, paste0("YR", 20:29)])
R> # 30 interaction dummies as instrumental variables
R> Z <- do.call(apply(QOB, 2, FUN = function(quarter) 
+    create_interactions(quarter, YOB)[[2]], simplify = F),
+    what = cbind)
R> # 9 year of birth dummies
R> YOB <- YOB[, -1]
R> # outcome
R> Y <- df$LWKLYWGE
R> # treatment
R> D <- df$EDUC
R> # baseline covariates
R> X <- cbind(YOB, df$RACE, df$MARRIED, df$SMSA, df$NEWENG, df$MIDATL,
+    df$ENOCENT, df$WNOCENT, df$SOATL, df$ESOCENT, df$WSOCENT, df$MT)
R> colnames(X) <- c(paste0("YR", 21:29), "race", "married", "smsa", "neweng", 
+    "midatl", "enocent", "wnocent", "soatl", "esocent", "wsocent", "mt")
\end{Sinput}
\end{Schunk}

Note that \code{Z} and \code{YOB} are linearly dependent.
Thus, we can drop the first $9$ columns of \code{X} for the treatment model, but they must be
present in the outcome model. Indeed, the instruments should only be interactions between the birth year and quarter but no main effects. We use \code{tsci\_forest} to assess the validity of the instruments. To do this, we consider a violation space sequence whose first element consists of the baseline covariates only, corresponding to valid instruments, and the second element the instruments themselves, corresponding to testing for a linear violation. We consider $1$ split of the data (\code{nsplits}) and $100$ trees (\code{num\_trees}) whose minimal node size is $5$ (\code{min\_node\_size}). 
We consider standard error estimates that come from a bootstrap approximation, which is the default  (\code{sd\_boot = TRUE}). 

\begin{Schunk}
\begin{Sinput}
R> set.seed(10)
R> vio_space <- list(Z) 
R> fit_angrist_tsci <- tsci_forest(Y = Y, D = D, Z = Z, 
+    X = X[, -seq_len(9)], W = X, vio_space = vio_space, nsplits = 1, 
+    min_node_size = 5, num_trees = 100)
\end{Sinput}
\end{Schunk}

The extended summary output below suggests that the instruments are actually invalid because the larger violation space is chosen (\code{q1}), also represented by the \texttt{1} under \texttt{invalid}. By default, the comparison method is used to choose a violation space. 
With this violation space, the treatment effect estimator is \texttt{0.14669} with a bootstrap standard error of \texttt{0.00389}. Consequently, it is significantly different from $0$ on the $5\%$ level.  

\begin{Schunk}
\begin{Sinput}
R> summary(fit_angrist_tsci, extended_output = TRUE)
\end{Sinput}
\begin{Soutput}
Statistics about the data splitting procedure:
Sample size A1: 164799 
Sample size A2: 82400 
Number of data splits: 1 
Aggregation method: DML 

Statistics about the validity of the instrument(s):
       valid      invalid non_testable 
           0            1            0 

Treatment effect estimate of selected violation space candidate(s):
              Estimate Std_Error    2.5%   97.5% Pr(>|t|)
TSCI-Estimate  0.14669   0.00389 0.13906 0.15432        0
Selection method: comparison 

Treatment effect estimates of all violation space candidates:
        Estimate Std_Error    2.5%   97.5% Pr(>|t|)
TSCI-q0  0.14402   0.00363 0.13690 0.15114        0
TSCI-q1  0.14669   0.00389 0.13906 0.15432        0

Statistics about the treatment model:
Estimation method: Random Forest 

Statistics about the violation space selection:
   q_comp q_cons Qmax
q0      0      0    0
q1      1      1    1

Statistics about the IV strength:
   IV_Strength IV_Threshold
q0     2282.93           40
q1     2134.61           40
\end{Soutput}
\end{Schunk}

\section{Practical guidance}\label{sect:guide}

In this section, we summarize the differences between \code{tsci\_forest}, \code{tsci\_boosting} and \code{tsci\_poly}. 
The random forest and boosting machine learning approaches may be employed if little is known about $f$. The polynomial approach, \code{tsci\_poly}, may be employed if the part of $f$ that is orthogonal to the violation is believed to be captured well by polynomials. With the machine learning approaches, more care is required to choose a violation space sequence. With the polynomial approach, the polynomial basis can be used to specify the violation space candidates.

\section{Summary and discussion} \label{sect:summary}

The primary goal of \pkg{TSCI} is to provide a user-friendly implementation of two stage curvature identification (TSCI)~\citep{GuoBuehlmann2022}. 
The TSCI method fills an important gap in the instrumental variable (IV) regression literature by providing a data-driven approach to test for invalidity of IVs and providing an effect size estimator that is robust to such violations. In particular, all instruments may be invalid.
In contrast to existing approaches for invalid IVs, TSCI only makes the very mild assumption that the treatment model and the IV violation are of a different functional form. Machine learning is employed to fit the treatment model, which allows us to capture complex nonlinearities and interactions.  
Nevertheless, \pkg{TSCI} should not be treated as a blackbox algorithm if machine learning is used because expert knowledge about the potential functional forms of the violation is required.

\section*{Computational details}

All packages used are available from the Comprehensive
\proglang{R} Archive Network (CRAN) at
\url{https://CRAN.R-project.org/}.
The results in Section~\ref{sect:using-tsci-ex2} were obtained using \proglang{R}~4.1.3. 
The random forest implementation in \pkg{ranger}~{0.13.3}~\citep{ranger}, the boosting implementation in \pkg{xgboost}~{1.6.0.1}~\citep{xgboost} and the fast matrix multiplication implementation in \pkg{Rfast}~{2.0.6}~\citep{Rfast} are used in \pkg{TSCI}~{2.0.1}.
All other results in this paper were obtained using \proglang{R}~4.2.3. 
The random forest implementation in \pkg{ranger}~0.14.1 \citep{ranger}, the boosting implementation in \pkg{xgboost}~1.7.5.1~\citep{xgboost} and the fast matrix multiplication implementation in \pkg{Rfast}~2.0.7~\citep{Rfast} are used in \pkg{TSCI}~2.0.1.
Additionally, we used \pkg{fda}~6.0.5~\citep{fda} for generating B-spline bases in this paper. We used \pkg{MASS}~7.3.58.3~\citep{MASS} to generate multivariate Gaussian data.
The computational complexity to compute the hat matrix using \code{tsci\_forest} is quadratic in the number of observations, whereas it is cubic with \code{tsci\_boosting}. Thus, using \code{tsci\_boosting} might be significantly slower than \code{tsci\_forest} for large datasets.

\section*{Acknowledgments}

We would like to thank Wei Yuan for providing an initial implementation of the random forest code that was developed when Wei Yuan was a research assistant at Rutgers University with ZG. We would like to thank Mengchu Zheng, currently research assistant at Rutgers University with ZG, for helpful comments. 
CE and PB received funding from the European Research Council (ERC) under the European Union’s Horizon 2020 research and innovation programme (grant agreement No. 786461). The research of ZG was supported in part by the NSF-DMS 1811857, 2015373 and NIH-1R01GM140463-01; ZG also acknowledges financial support for visiting the Institute of Mathematical Research (FIM) at ETH Zurich.

\bibliography{refs}


\begin{appendix}

\section{Encoding IV violation in $h$}\label{sect:appendix-h}

Subsequently, we argue that if the IV $\Zi$ is invalid, then $h\not\equiv 0$. Our argument illustrates that if $g$ does not depend on $\Zi$, but $\Zi$ is associated with the unmeasured confounders, this can again be represented as a model where $g$ does depend on the IV. 
If $g$ depends on $\Zi$, in which case the 
assumptions~\ref{assumpt:A2} or~\ref{assumpt:A3} are violated, the claim follows. Thus, let us consider an IV $\Zi$ that violates~\ref{assumpt:A2}. 
Consequently, we have $\Cov(f(\Zi, \Xi), \epsi) \neq 0$. 
Let us define $G(\Zi) = \E[\epsi | \Zi]$.  
Due to the exogeneity of $\Xi$, we have  $\Cov(f(\Zi, \Xi), \epsi - G(\Zi)) = 0$ and consequently $\Cov(f(\Zi), G(\Zi))\neq 0$. 
If we redefine $\epsi$ as $\epsi - G(\Zi)$, we have again an error that is not correlated with the instrument, but a $g$ that depends on $\Zi$, namely $G$.

\section{Bootstrap approach for standard deviation}\label{sect:appendix-b}

To simplify notation, we define
\begin{equation}\label{eq:def-M}
    \Mbold(\VboldAone) = \Omegabold^{\Tsf} \PVcompAone \Omegabold.
\end{equation}
The effect size estimator in~\eqref{eq:beta-estimator} can then be decomposed into
\begin{equation}\label{eq:beta-expansion2}
    \betahat - \beta = b(\VboldAone) + \frac{\DAonebold^{\Tsf} \Mbold(\VboldAone) \epsilon}{\DAonebold^{\Tsf} \Mbold(\VboldAone) \DAonebold} - \frac{\sum_{i = 1}^{n_1} [\Mbold(\VboldAone)]_{ii} \deltahatVAone_{i} \epshatVAone_{i}}{\DAonebold^{\Tsf} \Mbold(\VboldAone) \DAonebold},
\end{equation}
where $b(\VboldAone)$ is the sum of the second and third term in~\eqref{eq:beta-expansion} and may be small if a suitable candidate for the violation space is chosen, and the third term in~\eqref{eq:beta-expansion2} is the explicit form of the bias correction term in~\eqref{eq:beta-estimator}, where $\deltahatVAone = \DAonebold - \widehat{f}$ and 
\begin{displaymath}
    \epshatVAone = \PVWcompAone \bigg(Y - D \frac{\YAonebold^{\Tsf} \Mbold(\VboldAone) \DAonebold}
    {\DAonebold^{\Tsf} \Mbold(\VboldAone) \DAonebold}\bigg).
\end{displaymath}
The additional variance introduced by the bias correction term is asymptotically negligible under some regularity conditions~\citep{GuoBuehlmann2022}. However, in the finite sample setting, accounting for it might improve the accuracy of a standard error estimator.
Our bootstrap approach to derive such an estimator is as follows: 
for $1\leq i\leq n_1$, we center the residuals of the treatment and outcome model, $\widetilde{\delta}_i=\widehat{\delta}_i-\bar{\mu}_{\delta}$ with $\bar{\mu}_{\delta}=\frac{1}{n_1}\sum_{i=1}^{n_1}\widehat{\delta}_i$ and $\widetilde{\epsilon}_i=[\widehat{\epsilon}(V_{Q_{\max}})]_i-\bar{\mu}_{\epsilon}$
with $\bar{\mu}_{\epsilon}=\frac{1}{n_1}\sum_{i=1}^{n_1}[\widehat{\epsilon}(V_{Q_{\max}})]_i$, respectively. 
For some natural number $L$ and $1\leq l\leq  L,$ we then generate $\delta^{[l]}_i=U^{[l]}_i\cdot \widetilde{\delta}_i$ and $\epsilon^{[l]}_i=U^{[l]}_i\cdot \widetilde{\epsilon}_i$ for $1\leq i\leq n_1$, where $\{U^{[l]}_i\}_{1\leq i\leq n_1}$ are i.i.d. standard normal random variables.
Next, 
we compute the bootstrap analog of the second and third term in~\eqref{eq:beta-expansion2} 
according to
\begin{displaymath}
    N^{(l)}= \frac{\DAonebold^{\Tsf} \Mbold(\VboldAone) {\epsilon}^{[l]} - \sum_{i = 1}^{n_1}({[\Mbold(\VboldAone)]_{ii} \delta}^{[l]}_i \epsilon_i^{[l]} )}{{{D}_{\mathcal{A}_1}^{\Tsf} \Mbold(\VboldAone){D}_{\mathcal{A}_1}}}.
\end{displaymath}
Then, we obtain the empirical standard error of our effect size estimator from the bootstrap statistics $\{N^{(l)}\}_{1\leq l\leq L}$, and we denote it by ${\rm SE}_{\rm boot}$. Subsequently, a two-sided $95\%$ confidence interval is given by 
\begin{displaymath}
    (\widehat{\beta}-1.96\cdot{\rm SE}_{\rm boot},\widehat{\beta}+1.96\cdot{\rm SE}_{\rm boot}).
\end{displaymath}

\end{appendix}

\end{document}